\documentclass[%
 reprint,
 amsmath,amssymb,
 aps,superscriptaddress
]{revtex4-1}

\usepackage{graphicx}       
\usepackage{hyperref}       
\usepackage{physics}
\usepackage[utf8]{inputenc} 

\begin{document}

\title{Entanglement through qubit motion and the dynamical Casimir effect}
\author{Andr\'es Agust\'i­}
\affiliation{Instituto de F\'isica Fundamental, CSIC,
Serrano, 113-bis,
28006 Madrid (Spain)}
\author{Enrique Solano}
\affiliation{Department of Physical Chemistry, University of the Basque Country UPV/EHU, Apartado 644, 48080 Bilbao, Spain}
\affiliation{IKERBASQUE, Basque Foundation for Science, Maria Diaz de Haro 3, 48013 Bilbao, Spain}
\affiliation{Department of Physics, Shanghai University, 200444 Shanghai, China}
\author{Carlos Sab\'in}
\affiliation{Instituto de F\'isica Fundamental, CSIC,
Serrano, 113-bis,
28006 Madrid (Spain)}
\date{\today}
\begin{abstract}
We explore the interplay between acceleration radiation and the dynamical Casimir effect (DCE) in the field of superconducting quantum technologies, analyzing the generation of entanglement between two qubits by means of the DCE in several states of qubit motion. We show that the correlated absorption and emission of photons is crucial for entanglement, which in some cases can be linked to the notion of simultaneity in special relativity.
\end{abstract}
\maketitle

\section{\label{sec:level1} Introduction}

Superconducting quantum technologies together with circuit quantum electrodynamics (cQED) \cite{cQED1} form one of the most promising candidates for processing quantum information as well as the experimentation on the foundations of quantum mechanics. The advantages of this technology are, on the one hand, the strong coupling between the superconducting qubits and the resonant cavities, as well as the ability to widely tune this coupling and all the parameters of the system, allowing the investigation of new phenomena \cite{ArtificialAtomFluorescence, JCBreakdown} such as, among many others, the Dynamical Casimir Effect (DCE) \cite{ObservationDCE}  or the Unruh effect \cite{UnruhColloquium}.

The DCE is a member of a large family of effects linked to the quantum fluctuations of the vacuum, among which are the Lamb displacement \cite{Lamb}, the magnetic moment of the electron or the Casimir (static) effect \cite{Casimir1, ObservationCasimir1,ObservationCasimir2}. The latter is produced by a reduction in the density of modes imposed by certain boundary conditions, which leads to a pressure of radiation exerted by the vacuum. Its dynamical counterpart shares a similar origin, except for the fact that boundary conditions must be time-dependent, which can be achieved  by means of  SQUIDs \cite{ObservationDCE} . This effect has also been measured by modulating the effective speed of light \cite{ObservationDCE2}. On the other hand, the DCE can be considered a resource to generate quantum correlations, including quantum entanglement \cite{DCEentangles2,DCEentangles3,DCEentangles, discord, coherence}. In this paper, we will focus on this feature.

The Unruh effect is another member of the family of aforementioned phenomena, since it consists in the measurement of thermal radiation by a detector moving with constant proper acceleration through the quantum vacuum \cite{DaviesEffect,UnruhEffect}. Unfortunately, this has never been observed since it requires unreachable accelerations to generate detectable signals, although it can be increased by several orders of magnitude when two-level systems are accelerated through resonant cavities by means of non-adiabatic boundary conditions \cite{Scully}. However, this modification of the original effect is still difficult to reach experimentally and has not been observed. In this report we will analyse another scenario: it is possible to simulate the movement of a qubit in a cavity by modulating the qubit-cavity coupling in the same way said movement would \cite{SimulatedUnruh}, leading in general to the excitation of qubit and cavity from vacuum as in both variants of the effect. In the same way as with the DCE, this simulation of the Unruh effect can be used as a mechanism for the generation of entanglement \cite{UnruhEntangles}, only this time between a qubit and the photons of a cavity, not just in the field as in the first effect.

In this paper we consider an scenario where both the DCE and the simulated enhanced Unruh radiation can take place (see Fig. \ref{circuit}). It consists of two superconducting resonators sharing a common SQUID and each of them coupled to a superconducting qubit. DCE radiation can be generated by means of the modulation of the magnetic flux threading the SQUID, while the coupling of the qubits can be tuned to simulate their motion. We show that the correlated absorption and re-emission of the DCE radiation by the qubits is crucial for the generation of entanglement, as in the case where the qubits are static \cite{DCEentangles}. If the motion of the qubit preserves the correlated nature of the absorptions and emissions, entanglement is preserved as well. However, in general, uncorrelated motion of the qubits will result in the vanishing of entanglement, even for low simulated velocities. In the case of equal-length cavities, this physics can be linked to the notion of simultaneity in special relativity: breaking down simultaneity in the absorption would make entanglement vanish.

The rest of the paper has been structured as follows. First, we present the superconducting setup, its Hamiltonian and some relevant features. Then in the next section, we discuss the results of, on the one hand, applying perturbation theory to the calculation of the concurrence, and on the other hand, calculating this same magnitude solving the master equation of the system numerically. Finally, we summarize our conclusions.

\section{\label{sec:level1} Setup: DCE and simulated qubit motion}

\begin{figure}
 \centering
 \includegraphics[width=.49\textwidth]{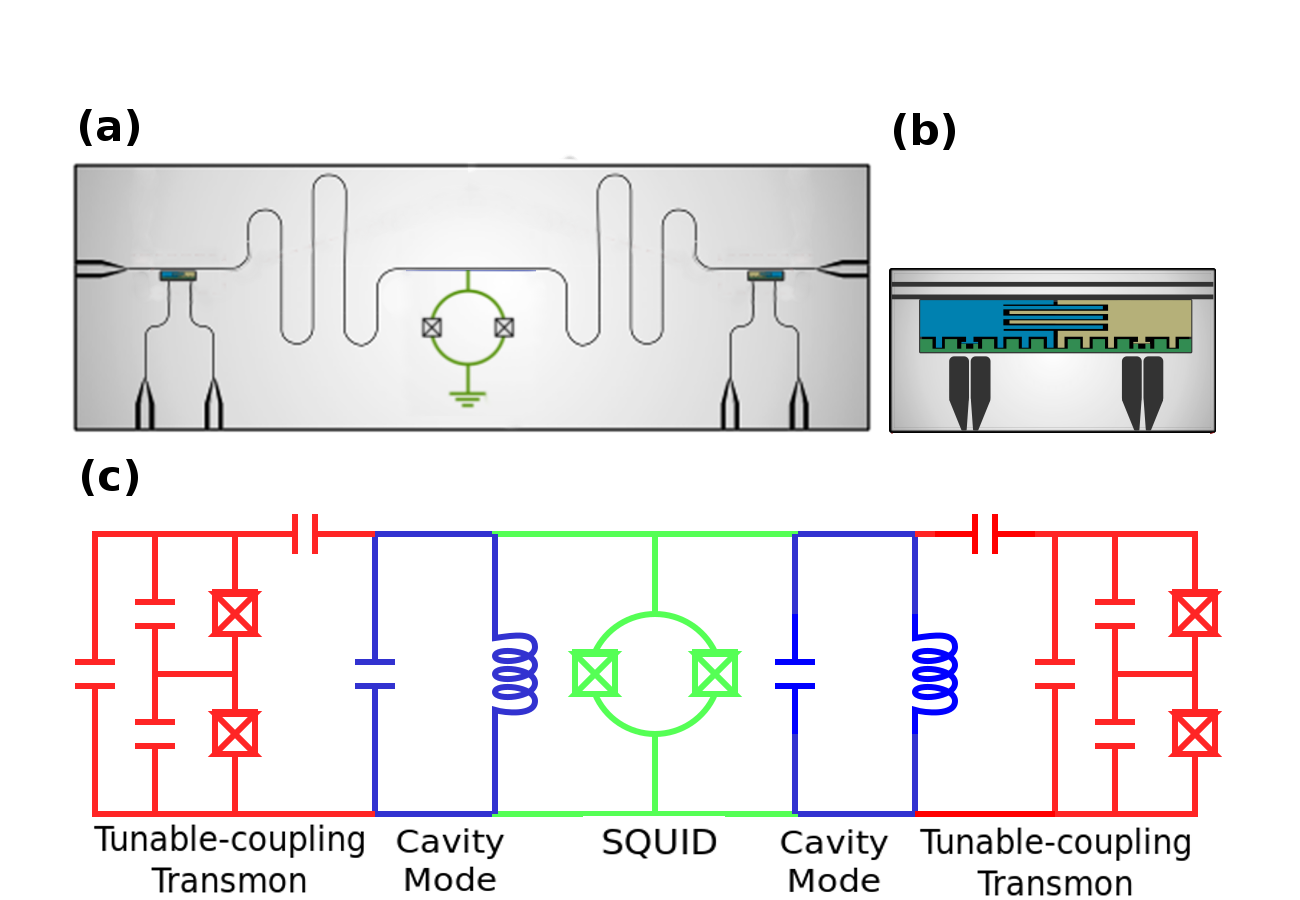}
 \caption{(a) Scheme of the setup consisting of two qubits, two resonators and a SQUID. (b) Scheme of the tunable-coupling transmons. (c) Scheme of the circuit, in red, the qubits, the relevant parameters being their characteristic frequencies $\omega_{q_{1,2}}$ and their coupling to the resonators $g_{1,2}(t)$. In blue are indicated the transmission line resonators or cavities, of which only one mode is considered. Their fundamental parameters are their characteristic frequencies $\omega_{c_{1,2}}$ as well as the coupling between them $g_{12}(t)$ due to the SQUID that couples them, in green. The capacitive coupling between cavities and qubits mentioned above is also indicated in red.
}
 \label{circuit}
\end{figure}

We consider a system composed of two superconducting qubits -in particular, modifications of the usual design of a transmon qubit \cite{TunableTransmon}- whose coupling to the electromagnetic field can be controlled by the magnetic flux threading the SQUIDs that compose them, which offers an opportunity to simulate the generation of acceleration radiation in the cavity-enhanced Unruh effect.  In addition, each of these qubits will interact directly with only one of the modes of only one of the two cavities or transmission line resonators of the system. These resonators interact in turn with each other by means of another SQUID, which allows to produce time-dependent boundary conditions in the resonators, giving rise to the DCE. An outline of this system can be found in Fig \ref{circuit}.

The Hamiltonian is
\begin{align}
 H &= \hbar\sum_{i=1}^2\left[
      \omega_{c_i}\left( a^\dagger_ia_i+\frac{1}{2} \right) + 
      \frac{\omega_{q_i}}{2} \sigma^z_i\right] +  \nonumber\\
      & + \hbar\sum_{i=1}^2g_i \cos(k_ix_{q_i}(t)) \sigma^x_i (a^\dagger_i + a_i) +\nonumber \\
      & + \hbar g_{1,2}(t)(a^\dagger_1 + a_1)(a^\dagger_2 + a_2),
\end{align}
where the sum runs over both cavities and both qubits. The first line contains the static Hamiltonian, where  $\omega_{c_i}$ is the frequency of cavity $1$ or $2$, depending on the subscript, as well as  $\omega_{q_i}$ is that of
the qubits. On the other hand, $a^\dagger_i$ and $a_i$ are the creation and annihilation operators of the corresponding cavity and $\sigma^z_i$ the third Pauli operator of each qubit. The second line of the equation contains the interaction of the qubits with their cavities, being $k_i$ the wave vector of the cavity, $g_i$ the maximum intensity of its coupling,  $\sigma^x_i$ the first Pauli operator of the qubit and $x_{q_i}(t)$ simulated trajectory of qubit motion. Experimentally, the product $k_ix_{q_i}(t)$  is actually $f = \phi(t)/\phi_0$ with $\phi_0$ the quantum of magnetic flux and $\phi(t)$ the flux through the SQUID \cite{SimulatedUnruh,UnruhEntangles} that may be controlled from the outside with a typical nanosecond resolution. The third line contains the interaction between the cavities, being $g_{1,2}(t)$ the time-dependent coupling assuming that the boundary conditions produced by the SQUID do not destroy the structure of normal field modes or make resonances of new modes with the qubits. This is the case when $g_{1,2}(t) = g_0\cos(\omega_d t)$ with $\omega_d = \omega_{c_1} + \omega_{c_2}$ matching the sum of the cavity mode frequencies. Moreover, in this case the interaction Hamiltonian can be approximated by a two-mode squeezing Hamiltonian \cite{DCEentangles}:
\begin{align}
      g_{1,2}(t)(a^\dagger_1 + a_1)(a^\dagger_2 + a_2) \rightarrow 
      g_0 /2(a^\dagger_1a^\dagger_2 + a_1a_2).
\end{align}

So, finally:

\begin{align}  
 \label{H}
 H &= \hbar\sum_{i=1}^2\left[
  \omega_{c_i}\left( a^\dagger_ia_i+\frac{1}{2} \right) + 
  \frac{\omega_{q_i}}{2} \sigma^z_i\right] + \nonumber \\
  & + \hbar\sum_{i=1}^2g_i \cos(k_ix_{q_i}(t)) \sigma^x_i (a^\dagger_i + a_i) + \nonumber\\
  & + \hbar g_0 /2
  (a^\dagger_1a^\dagger_2e^{i \omega_d t} + a_1a_2e^{- i \omega_d t}),
\end{align}
  
Previous work \cite{SimulatedUnruh} has shown that, even for moderate values of the coupling, the modulation of the coupling strength might resonate with the counterrotating terms of the Hamiltonian. Therefore, we will not perform the rotating wave approximation (RWA).

Finally, notice that if either or both of the trajectories $x_i(t)$ are changed by $L_i - x_i(t)$ and the relevant coupling constant $g_i \rightarrow - g_i$ is inverted, then the Hamiltonian does not undergo any change, which is quickly deduced using the expression of the cosine of the sum and substituting $k_i = \pi/L_i$. This symmetry can be interpreted as a mechanism by which a path $x_{q_i}(t)$ that passes between the two ends of the cavity in a finite time can be extended beyond that time reflecting it with respect to the center of the cavity. That is, if for a time $\tau$  we have $x_i(\tau) = L_i$  and at a later time $x(\tau+\delta t)$ exceeds $L_i$, then the path can be modified as follows $x_i'(\tau + \delta t) = L_i - x_i(\delta t)$ , producing the same Hamiltonian which governed the evolution up to $\tau$, except for the sign of $g_i$. In other words, this symmetry offers a natural bounce condition to continue trajectories that reach the ends of the cavities, so natural that it will prove useful throughout the work.

\section{\label{sec:level1} Results}

We will use two different methods to analyze entanglement generation in the setup discussed in the previous section, namely perturbation theory to obtain an approximate expression of the concurrence and then numerical simulations to integrate the master equation governing the system by means of the Python package QuTiP \cite{QuTiP}. These two methods make up the following two subsections. As a final section, dissipation and temperature are addressed.

\subsection{\label{sec:level2} Perturbative results}

The global state $\ket{\Psi(t)}$ under Hamiltonian (\ref{H}) is expanded up to third order in the couplings $g_1$, $g_2$, $g_0$ from the ground state $\ket{00gg}$, where $\ket{0}$, $\ket{1}$ and so on are the Fock number states of each cavity and  $\ket{g}$, $\ket{e}$ are the ground and excited states of each qubit. We aim to compute the entanglement dynamics of the qubits, so it is necessary to trace over the cavity field states, which leads to the following density matrix:

\begin{equation}
 \rho^{(3)}_{qubits} = 
 \begin{pmatrix}
  0 & 0 & 0 & \rho_{14} \\
  0 & 0 & 0 & 0 \\
  0 & 0 & 0 & 0 \\
  \rho_{41} & 0 & 0 & 1
 \end{pmatrix}
 \label{state}
\end{equation}
where $\rho_{14}$ the matrix element $\bra{ee}\rho\ket{gg}$ and the not zero entry in the diagonal corresponds to $\bra{gg}\rho\ket{gg}$. For a more detailed derivation refer to Appendix A. Then, using the concurrence $C(\rho)$ \cite{Concurrence} as entanglement measurement, we find:
 \begin{align} \label{concurrence}
        &C(\rho) = 2|\rho_{14}| 
        = 2|\bra{ee}\rho\ket{gg}| 
        = 2|\bra{00ee}\rho_{total}\ket{00gg}| \nonumber\\
        &= g_1g_2g_0 \times\nonumber\\
        &
        \Biggl|\int_0^tdt_1\int_0^{t_1}dt_2\int_0^{t_2}dt_3
        \cos(f_1(t_1))\cos(f_2(t_2))e^{i\omega_dt_3} + \nonumber\\
        &+
        \int_0^tdt_1\int_0^{t_1}dt_2\int_0^{t_2}dt_3
        \cos(f_1(t_2))\cos(f_2(t_1))e^{i\omega_dt_3}
        \Biggr|,
    \end{align}
   where $f_i=k_ix_i(t)$. It should be noted that in this system the concurrence can be interpreted in a very intuitive fashion: it is proportional to the probability that two photons are emitted, one in each cavity, and that each qubit absorbs one. At first sight, Eq. (\ref{concurrence}) seems to conclude that the qubit motion will only reduce concurrence, since integrating cosines will not provide any further contribution. However, these can cause a resonance, either with each other or with the term due to the emission. In the following paragraphs, we will study these resonances produced by different trajectories, paying attention to the conditions that must be fulfilled for their existence.

\subsubsection{\label{sec:level3} Static qubits}
If the qubits are static Eq. (\ref{concurrence}) further simplifies, since the cosine functions are all constant. By performing the first integration, we get
$\int_0^{t_2}dt_3 e^{i\omega_d t_3} = (e^{i\omega_d t_2}-1)/i\omega_d$ 
 and the constant term will eventually give rise to:
\begin{align}
 C_{\text{rest}} = \frac{g_0g_1g_2}{\omega_d}t^2 + O(t).
\end{align}
This quadratic behaviour of the concurrence seems to be in agreement with \cite{DCEentangles} for the moderate values of time where the perturbative approach is valid -the perturbative approximation will eventually break down, as we will see in detail below.

\subsubsection{\label{sec:level3} Constant velocity}

Particularizing Eq. (\ref{concurrence}) for the case of qubits moving with constant velocities $v_i$ with initial positions at $x=0$, we get :
\begin{align}\label{concurrencev}
    C_{\text{v const}}
    &= \frac{g_1g_2g_0}{4\omega_d} \times\nonumber\\
    &
    \Biggl|\int_0^tdt_1\int_0^{t_1}dt_2
    (e^{ik_1v_1t_1}+e^{-ik_1v_1t_1}) \times\nonumber \\
    &(e^{ik_2v_2t_2}+e^{-ik_2v_2t_2})(e^{i\omega_dt_2}-1) + \nonumber\\
    &+
    \int_0^tdt_1\int_0^{t_1}dt_2
    (e^{ik_1v_1t_2}+e^{-ik_1v_1t_2}) \times\nonumber \\
    &(e^{ik_2v_2t_1}+e^{-ik_2v_2t_1})(e^{i\omega_dt_2}-1)
    \Biggr|
\end{align}
By inspection of Eq. (\ref{concurrencev}), we find that resonances will appear if either or both following conditions are verified:
\begin{align}\label{conditions}
 \omega_d &= k_i|v_i| \nonumber\\
 k_1|v_1| &= k_2|v_2|
\end{align}
Note that $\omega_d = \omega_1 + \omega_2$, so the above conditions turn into:
\begin{align}\label{conditions2}
    c/L_1 + c/L_2 &= |v_i|/L_i \nonumber\\
    |v_1|/L_1 &= |v_2|/L_2,
\end{align}
The first one implies a superluminal velocity of at least one of the qubits, and was already found in \cite{lastcite}. It is related with the emission of Ginzburg radiation at superluminal constant velocities. The second one -since both qubits start in the same position- means that the distance of the qubits to $x=0$ -in units of the corresponding cavity length- is always the same for both qubits  $x_1/L_1 = -x_2/L_2$. Then the hamiltonians of both qubits are equivalent at any time, entailing that the absorption and re-emission of DCE photons is perfectly correlated. In particular, if the lengths of the cavities are equal $L_1=L_2$ then the distances are exactly the same, which means that absorptions and emissions occur always simultaneously. This suggests an interesting link between the generation of entanglement in a quantum setup and a key notion of special relativity, such as simultaneity.

When both conditions in Eq. (\ref{conditions2}) are met at the same time, namely $|v_1|/L_1 = |v_2|/L_2 = c/L_1 + c/L_2$, then we find:
\begin{align}\label{concurrencev1}
C_{\text{v const, 1}} = \frac{g_0g_1g_2}{\omega_d^2}|\sin(\omega_dt)|t + O(t^0),
\end{align}
while if only the first condition is met $|v_1|/L_1 = c/L_1 + c/L_2 \neq |v_2|/L_2 = kv/\pi$, then
\begin{align}\label{concurrencev2}
C_{\text{v const, 2}} = \frac{g_0g_1g_2}{2\omega_dkv}|\sin(kvt)|t + O(t^0).
\end{align}
In the latter case, it is the velocity of the non-resonant qubit and not the mirror frequency what modulates the generation of concurrence. Even if we assume $kv \approx \omega_d$, then the concurrence in Eq. (\ref{concurrencev2}) is reduced by a factor 1/2 with respect to the concurrence in Eq.(\ref{concurrencev1}), which highlights again the importance of the correlations among the absorptions and emissions of photons. 

Interestingly, we can use the symmetry of the Hamiltonian described at the end of Section \ref{sec:level1}, that in this type of trajectory translates into an inversion of the speed of the qubits when they try to leave the cavities. The full symmetry would change the sign of the relevant coupling, but these signs can be canceled if both trajectories arrive at the ends of the cavities simultaneously and their speeds are then inverted. With these bounces, trajectories of constant velocity can be extended in time. The concurrence (\ref{concurrencev}) inherits this Hamiltonian symmetry: after $n$ bounces the generated entanglement is $n$ times the entanglement after the first bounce. We see that in this case the simultaneity in the bounce plays a crucial role by simplifying the calculation of the concurrence and extending their relevance.
Finally, if the second condition in Eq. (\ref{conditions2}) is met but not the first it is not possible to obtain a closed analytical expression of the concurrrence. However, we will present numerical results in the next section.

\subsubsection{\label{sec:level3} Other trajectories}

\begin{figure}
    \includegraphics[scale=.45]{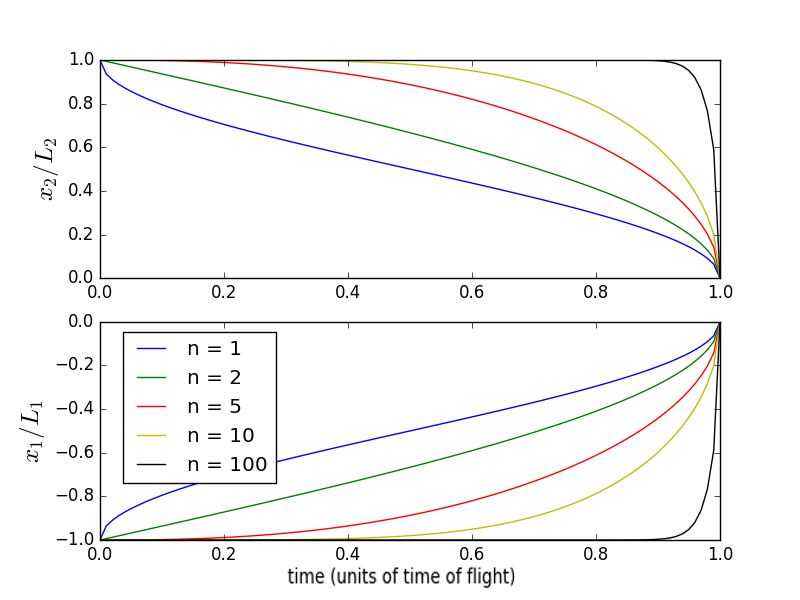}
    \caption{Trajectories of the qubits, in units of the lengths of their cavities and of the total flight time, for equation (\ref{eq_arccos}) and $n = 1,2,5,10,100$, from higher to lower in the $x_1/L_1$ subplot.}
    \label{fig_arccos}
\end{figure}

A convenient family of qubit trajectories would be given by:
\begin{align}
    x_1(t) &= \frac{L_1}{\pi} \arccos 
    \left(2\left(\frac{t}{\tau}\right)^n - 1\right) \nonumber \\
    x_2(t) &= -\frac{L_2}{\pi} \arccos 
    \left(2\left(\frac{t}{\tau}\right)^n - 1\right),
    \label{eq_arccos}
\end{align}
$\tau$ being the flight time of the qubits, namely the time that it takes for each qubit to traverse its cavity. These trajectories can be seen in Fig. (\ref{fig_arccos}). They exhibit divergences in the velocity and the acceleration when time gets close to the flight time, and also at $t=0$ for the particular case $n=1$.

We find the following concurrence:
\begin{align}\label{concurrencearc}
    C_{\text{arccos}} = 
    \frac{4g_0g_1g_2}{\omega_d(n+1)^2}
    \frac{v^{2n}}{L^{2n}}t^{2n+2} + O(t^{2n}).
\end{align}
Therefore, with these trajectories we are able to produce resonances with arbitrary powers of time. Of course, all the above is restricted by the perturbative approximations adopted, which will eventually break down. We now proceed to show results obtained by a numerical integration of the master equation of the system, which enables the exploration of the long-time dynamics.

\subsection{\label{sec:level2} Numerical results}

\subsubsection{\label{sec:level21} Static qubits and constant velocities}

\begin{figure}
\centering
\includegraphics[scale=.45]{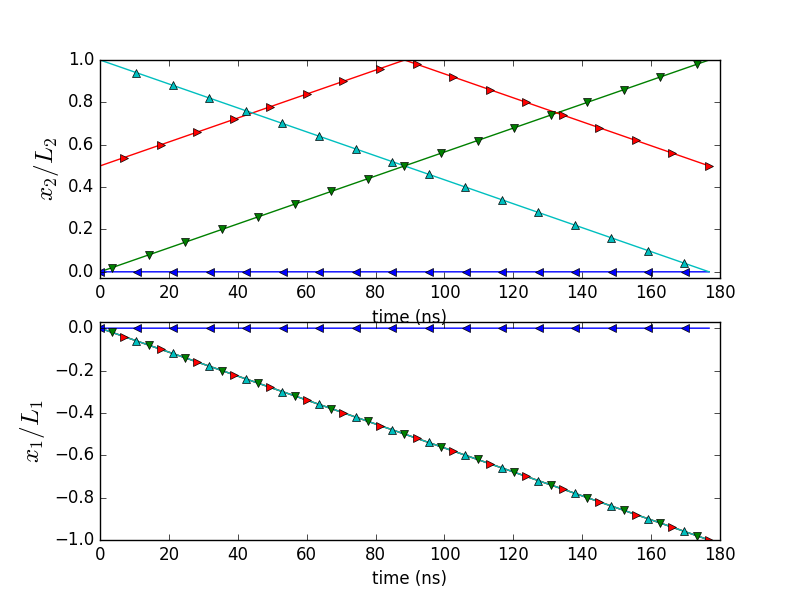}
\caption{Qubit trajectories in units of $L_i$. Dark blue left-facing triangles: static qubits. Green down-facing triangles: initial positions $x_i(0) = 0$ and opposite velocities $0.0001(\omega_1 + \omega_2)$. Red right-facing triangles: same velocities as in green but different initial positions $x_1(0)=0$ y $x_2(0)=L_2/2$. Cyan up-facing triangles: Same velocities but initial positions $x_1(0)=0$ y $x_2(0)=L_2$ }
\label{trayectories_num}
\end{figure}

We consider four trajectories (see Fig. (\ref{trayectories_num})). 
The first one is the case where the qubits are static, already discussed in \cite{DCEentangles}. Then, we consider the trajectory analyzed perturbatively in the previous section, where the qubits start at $x=0$ and move with opposite velocities, giving rise to correlated -simultaneous for equal cavities- absorption and emission of photons. The third case is related to the second by the symmetry relation discussed throughout this work, since one qubit starts at the other end of its cavity. Finally, we consider a trajectory which breaks the correlations among the absorptions and emissions of photons, since one qubit starts out at the center of its cavity while the other starts at 0. 

\begin{figure}
    \centering
    \includegraphics[scale=.45]{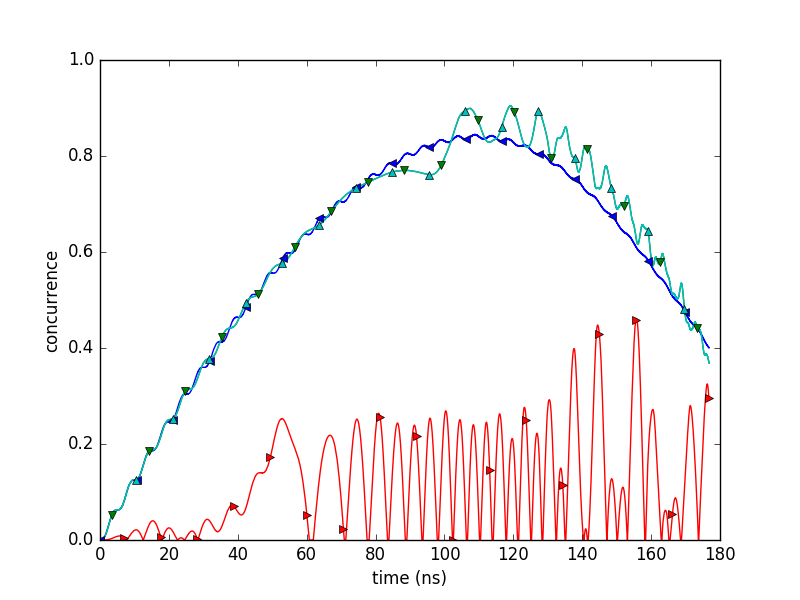}
    \caption{Concurrences for the trajectories of Fig. \ref{trayectories_num}. The rest of the relevant parameters are  $\omega_1/2\pi = 4\text{ GHz}$ y $\omega_2/2\pi = 5\text{ GHz}$ for the qubits and cavity modes, with coupling strengths $g_0 = 0.001 \omega_1$ y $g_1 = g_2 = 0.04 \omega_2$. The maximum values of the concurrence are  $0.844$, $0.904$, $0.461$ and $0.904$, attained at $108.4 \text{ ns}$, $119.0 \text{ ns}$, $155.6 \text{ ns}$ y $119.0 \text{ ns}$, respectively.}
    \label{concurrences_num}
\end{figure}

In Fig. \ref{concurrences_num} we show the numerical results for the concurrences at long times. We reproduce the results for the static case in \cite{DCEentangles} up to the point of maximum concurrence, where they propose to switch off the coupling in order to optimize the entanglement generation. We show that in the two trajectories which preserve the correlated absorptions and emissions the high concurrence is indeed preserved, even achieving larger maximum values. Finally, in the asymmetric case, the concurrence is significantly reduced, as expected. Interestingly, this effect occurs already at low non-relativistic velocities, which highlights the role of the correlation/simultaneity in the generation of entanglement in this setup.
In order to achieve the regime of large velocities, it is convenient to use trajectories similar to the ones in Section \ref{sec:level3}. We will explore them in the next subsection.

\subsubsection{\label{sec:level22} Other trajectories}

\begin{figure}
\centering
\includegraphics[scale=.45]{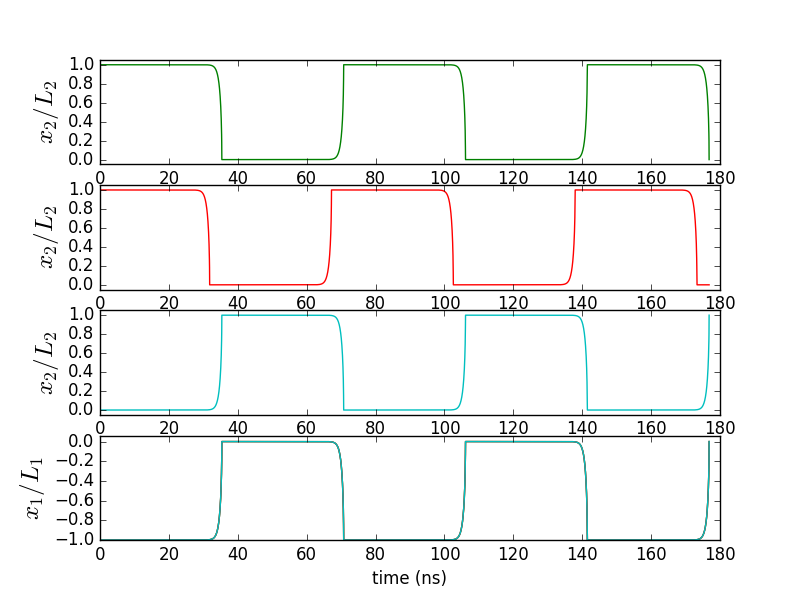}
\caption{Qubit trajectories. The bottom plot is $x_1(t)$ for all cases, as given in  (\ref{eq_arccos_num}) for $n=100$. In the top one, in green,  $x_2(t)$ given as well by  (\ref{eq_arccos_num}). In the second one, in red, $x_2(t+0.1\tau)$, while in the third one, cyan, $x_2(t+\tau/2)$.  }
\label{arc10_trayectories}
\end{figure}

In this case, we have considered the following trajectories (see Fig. \ref{arc10_trayectories}). We first use similar trajectories as in Eq. (\ref{eq_arccos}), but extended by means of the bounce symmetry:
\begin{align}
 f_{n}(x) &= \frac{1}{\pi} \arccos (2x^n-1) \nonumber \\
 x_1(t) &= -L_1
 f_{n}\left(\frac{t}{\tau} - \left \lfloor{\frac{t}{\tau}} \right\rfloor\right)
 & \text{if } \left \lfloor{\frac{t}{\tau}} \right\rfloor \text{even} 
 \nonumber \\
     &= -L_1 + L_1
 f_{n}\left(\frac{t}{\tau} - \left \lfloor{\frac{t}{\tau}} \right\rfloor\right) 
 & \text{if } \left \lfloor{\frac{t}{\tau}} \right\rfloor\text{odd}
 \nonumber \\
 x_2(t) &= L_2
 f_{n}\left(\frac{t}{\tau} - \left \lfloor{\frac{t}{\tau}} \right\rfloor\right) 
 & \text{if } \left \lfloor{\frac{t}{\tau}} \right\rfloor \text{even} 
 \nonumber \\
     &= L_2 - L_2
 f_{n}\left(\frac{t}{\tau} - \left \lfloor{\frac{t}{\tau}} \right\rfloor\right) 
 & \text{if} \left \lfloor{\frac{t}{\tau}} \right\rfloor \text{odd},
 \label{eq_arccos_num}
\end{align}
 $\tau = v_1/L_1 = v_2/L_2$ being again the time of flight which takes the qubits to traverse their cavities and $ \lfloor{x} \rfloor$ being the floor function. In this case, as it can be seen in Fig. \ref{arc10_trayectories}, the trajectories are synchronized in such a way that the absorption and emission of photons is correlated. 

\begin{figure}
\centering
\includegraphics[scale=.45]{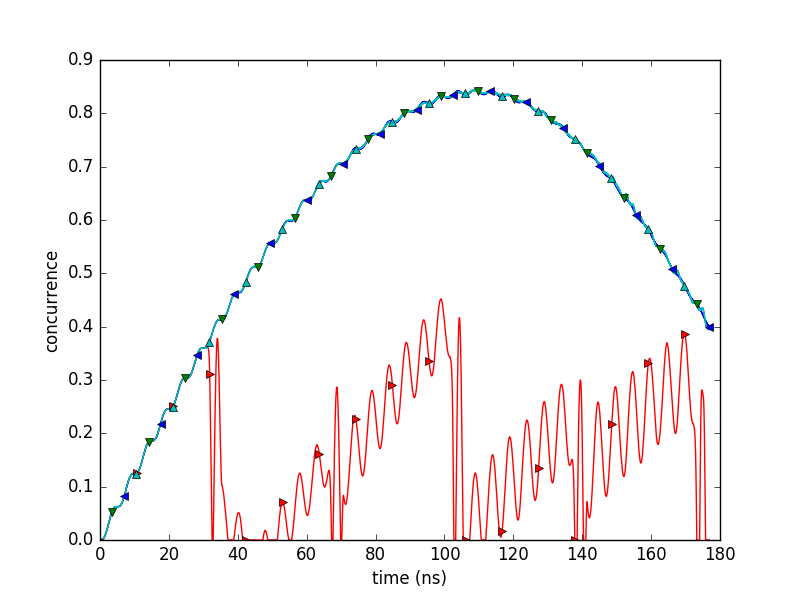}
\caption{Concurrences for the trajectories in Fig. \ref{arc10_trayectories}, using the same color code, but with down-facing triangle markers for green, right-facing for red and up-facing for cyan. As a reference, we plot in dark blue with left-facing triangles the static case.}
\label{arc10_concurrences}
\end{figure}

\begin{figure}
    \centering
    \includegraphics[scale=.45]{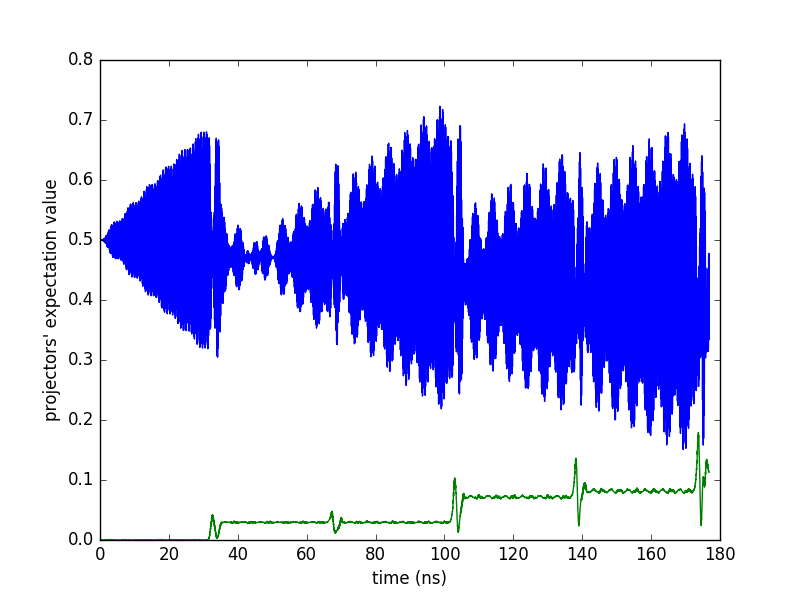}
    \caption{Expectation value of the Bell state projectors $\ket{\phi_{+}}$ (blue, upper plot) and $\ket{\psi_{+}}$ (green, lower). The other Bell states $\ket{\phi_{-}}$ and $\ket{\psi_{-}}$ have been omitted given that they take the same values, but with a nanosecond phase difference. Notice the remarkable similarity between the dynamics of the concurrence and the Bell state projector over $\ket{\psi_{+}}$ in this low-concurrence case. }
    \label{bell}
\end{figure}

Another case that we consider is obtained by replacing in Eq. (\ref{eq_arccos_num}) $x_2(t)$  by $x_2(t + \tau/2)$, which preserves the symmetry between the qubits. Finally, making instead the replacement with, for instance,  $x_2(t+0.1\tau)$, then the qubits are out of phase. We show the concurrences for these trajectories in Fig. \ref{arc10_concurrences}, which again shows that the correlations/simultaneity between the qubits are crucial to understand the magnitude of entanglement generation. In order to further illuminate this point, it is interesting to discuss the population in the Bell basis. In the static case and all the cases where the qubit positions are correlated in the way discussed above, photons are emitted and absorbed in pairs, and therefore it is expected that all the population is in the Bell states $\ket{\phi_\pm} = 1/\sqrt{2}(\ket{gg} \pm \ket{ee})$. However, if the positions of the qubits are uncorrelated it is possible that one of the qubits emits or absorbs a photon while the other does not. This enables the population of the other Bell states $\ket{\psi_{\pm}} = 1/\sqrt{2}(\ket{ge}\pm\ket{eg})$. In Fig. \ref{bell} we confirm that this is indeed the case: only in the low-concurrence case a significant population  eventually appears in $\ket{\psi_{\pm}}$. Comparing Fig. \ref{arc10_concurrences} with \ref{bell}, we see that jumps in the population of $\ket{\psi_{\pm}}$ are correlated with falls in the value of the concurrence, as expected.

\subsection{\label{sec:level2} Dissipation and Temperature}
As a final remark to this analysis, dissipation and temperature must be taken into account. The typical temperature in superconducting circuits experiments is of the order of $10 - 50$ mK, giving a thermal photon number of $10^{-9} - 10^{-2}$. This is clearly negligible and hasn't been considered in the calculations. Analogous simulations to those presented in this letter have been run considering relaxation and dephasing times up to $10^4$ ns and resonator decoherence times of $10^5$ ns, values at experimental reach by superconducting circuits technology \cite{devoret2013}, as are the values of the rest of parameters such as characteristic frequencies and couplings. This dissipation proved to be irrelevant in the dynamics of the system, as initially supposed.

\section{\label{sec:level4} Conclusions}

We have analyzed the entanglement dynamics between two qubits in a system where each one interacts with a resonant cavity with tunable coupling, which allows to simulate their motion. The cavities interact in turn with each other through a SQUID, which implements a boundary condition that can be modulated by the magnetic flow threading it. This results in a two-mode squeezing hamiltonian which is the source of entanglement. We show that a high degree of entanglement can be generated both in the case where the qubits are static -previously discussed in \cite{DCEentangles}- and in the cases where the motion preserves the fact that photons are absorbed and re-emitted in pairs -one by each qubit- populating only the Bell states $\ket{\phi_{\pm}}$. Otherwise, if the motion of the qubits is such that photons can be emitted or absorbed by only one qubit, we find that the states $\ket{\psi_{\pm}}$ are also populated and the concurrence is dramatically reduced. If the cavities are equal in length, this means that high-concurrence trajectories are characterized by simultaneous absorption and emission of photons, which suggests an interesting link with the notion of simultaneity in special relativity. 

These results pave the way for the exploration of special relativistic effects in a quantum setup. For instance, we envision the quantum simulation of the \textit{gedanken} textbook experiments where trains moving at relativistic speeds are used to illustrate the relativity of simultaneity. In our setup, the magnitude of entanglement could be used, in principle, as a witness of simultaneity, and vice versa. Our results are fully within reach of current technology. 

\section*{Acknowledgements}
A.A. and C. S have received financial support through the Junior Leader Postdoctoral Fellowship Programme from “la Caixa” Banking Foundation (LCF/BQ/LR18/11640005).
E.S. acknowledges support from the projects QMiCS (820505) and OpenSuperQ (820363) of the EU Flagship on Quantum Technologies, Spanish MINECO/FEDER FIS2015-69983-P and Basque Government IT986-16.

\section*{Appendix A}
As pointed out at the beginning of section III.A, the global state of both qubits and resonator modes is expanded up to third order in the couplings which gives non-zero projections onto the states:
$\{\ket{00gg}, \ket{11gg}, \ket{10ge}, \ket{01eg}, \ket{00ee}, \ket{22gg}, \ket{21ge},$ $\ket{12eg}, \ket{33gg}\}$. When performing the partial trace over the fields many of this states (those with the qubits in the same state) will mix, leading to a density matrix with the following expression:
\begin{equation}
 \rho_{qubits} = 
 \begin{pmatrix}
  \rho_{11} & 0 & 0 & \rho_{14} \\
  0 & \rho_{22} & 0 & 0 \\
  0 & 0 & \rho_{33} & 0 \\
  \rho_{41} & 0 & 0 & \rho_{44}
 \end{pmatrix}
\end{equation}
Where the notation for subscripts is the same as for equation (4). The expression of these matrix elements in terms of the perturbative coefficients of $\ket{\Psi}$ are:

\begin{align*}
\rho_{11} = 
\bra{ee}\rho\ket{ee} &= 
|c^{(3)}_{00ee}|^2 \\
\rho_{22} = 
\bra{eg}\rho\ket{eg} &= 
|c^{(2)}_{01eg}|^2 + |c^{(3)}_{12eg}|^2 \\
\rho_{33} = 
\bra{ge}\rho\ket{ge} &= 
|c^{(2)}_{10ge}|^2 + |c^{(3)}_{21ge}|^2 \\
\rho_{44} = 
\bra{gg}\rho\ket{gg} &= 
|c^{(0,2)}_{00gg}|^2 + |c^{(1,3)}_{11gg}|^2 + |c^{(3)}_{22gg}|^2 \\
\rho_{14} = 
\bra{ee}\rho\ket{gg} &= 
c^{(3)*}_{00ee}c^{(0,2)}_{00gg}
\end{align*}

Where the superscripts indicate the order in the perturbative expansion at which the coefficient appears -or in the case of multiple indices, the orders at which new contributions must be taken into account. Notice that since $\ket{\Psi}$ has been expanded up to third order, any term in the matrix elements with a greater power in the couplings (that is, adding their superscripts) must be considered negligible. With this in mind, the qubits' density matrix reads:
\begin{equation}
 \rho_{qubits} = 
 \begin{pmatrix}
  0 & 0 & 0 & c^{(3)*}_{00ee} \\
  0 & 0 & 0 & 0 \\
  0 & 0 & 0 & 0 \\
  c^{(3)}_{00ee} & 0 & 0 & 1
 \end{pmatrix}
\end{equation}
Which proves equation (4).

\end{document}